\documentclass[aip,jcp,twocolumn,superscriptaddress,floatfix,10pt]{revtex4-1}
\usepackage[pdfpagemode=UseNone,pdfstartview=FitH]{hyperref}
\usepackage{graphicx}
\usepackage{epstopdf}
\usepackage{mathptmx}
\usepackage{courier} 
\normalfont
\usepackage[T1]{fontenc}
\usepackage{amsmath,amssymb,amsthm}

\def\br{{\bf r}}
\def\bof{{\bf f}}
\def\delr{{\Delta \br}}
\def\nhat{{\bf\hat n}}
\def\xhat{{\bf\hat x}}

\def\Ohat{\hat O}
\def\px{\partial_x}
\def\half{\frac{1}{2}}

\begin{document}
\title{Photo-mechanical energy conversion using polymer brush dissociation}

\author{J. M. Deutsch}
\affiliation{Department of Physics, University of California, Santa Cruz CA 95064}

\begin{abstract}
A device is investigated that continuously and directly converts
light into mechanical energy, using polymers and photodissociation.
A polymer brush tethered to a surface, is brought into contact with a parallel
plate a small distance above it that contains reaction sites where
photodissociation of bound polymer and light can occur. Under the
appropriate conditions, the collective effect of these polymers is
to apply a force parallel to the plates, converting
incoming light into mechanical work. Numerical work is carried out
to understand this effect, a three dimensional Langevin simulation,
solution to the Fokker Planck equation, and a one dimensional Monte
Carlo simulation.  Theoretical analysis of the Fokker Planck equation
is used to study a model where equilibration of the unbound state
occurs and equilibration to a metastable equilibrium is achieved
in the bound state. It is shown that the work per cycle can be made
much larger than the thermal energy but at the expense of requiring
a greatly diminished photodissociation rate. Parameters
are discussed in order optimize mechanical energy conversion.
\end{abstract}
\maketitle

\section{Introduction}
\label{sec:Introduction}

There are many proposals to convert light into mechanical energy using smart
polymeric photo-responsive materials~\cite{IrieKunwatchakun,Irie,SuzukiHirasa,BehlLendlein,RoyGupta}
or the synthesis of individual molecules that lead to rotary or linear motion~\cite{credi2006}.  

Photoresponsive materials
use a conformational change of a macroscopic collection of polymers in response
to light or an external change in temperature and pH. This will cause a change
in volume. If this process is reversible, which it is in many instances, this
device can be cycled to derive mechanical energy~\cite{IrieKunwatchakun,Irie,SuzukiHirasa}. 

In general, there are many scenarios leading to a change in configuration of a molecule in
response to light~\cite{Irie}: {\em cis-trans} isomerization, zwitter ion formation, radical
formation, ionic dissociation, and ring formation/cleavage. If such
light-sensitive elements are incorporated into a macromolecular system, this can
in principle, through thermodynamic cycles, to conversion of light to
mechanical power.

Remarkable advances in the design of molecules have led to prototypes for
light-driven synthetic molecular motors~\cite{credi2006}. Many of these are
based on photoisomerization reactions that enable rotary motion of molecules
when combined with thermal rotation steps~\cite{terWiel2005,Ruangsupapichat,Vicario,credi2006}. 
It has also been possible to make threading-dethreading systems~\cite{Ashton1998,credi2006},
and shuttle molecular rings~\cite{Ashton2000,credi2006}. There has even been
progress on a macroscopic level, such as creating droplet motion by modifying surface
properties~\cite{Berna} or inducing mechanical deformation~\cite{Liu} similar to
what is achieved with photo-responsive materials~\cite{IrieKunwatchakun,Irie,SuzukiHirasa,BehlLendlein,RoyGupta}.

Much of the research in this area is inspired by biological machines such as myosin II that work by
quite a different principle than man-made macroscopic motors~\cite{BustamanteKellerOster}. Biological
motors use chemical energy rather than photons to produce mechanical power however this
difference does not affect the basic mechanism of operation. A myosin
head can be thought of as being in two states, bound or unbound. Thermal noise in the unbound state can cause
the head to bind to actin, producing a force. The hydrolysis of ATP releases energy
causing the head to return to the unbound state.  
The biochemistry of a real motor protein is considerably more complicated, but
by simplifying this description to one involving only these two states,
the motion can be analyzed, and 
it is easily seen that electromagnetic energy can be used instead of chemical
energy~\cite{ProstPRL}. 

In this paper, I investigate the use of photons in powering a two state motor
system similar to biological motors. I consider
creating motion between two surfaces that are very close together by placing
an asymmetric polymer brush between them.
This could potentially have advantages. A continuous source of light would create
a motion that is constant on a macroscopic scale, that could for example, rotate
the surfaces relative to each other. It would not require any additional
mechanisms to keep it moving, other than the microscopic motion of molecules
between the plates. 

The device proposed here falls into a distinct category different than the
experimental approaches above. It is not a macroscopic smart
material that changes properties in response to external stimuli. The device
described in the next section works on the scale of an individual polymers, 
and the force generated on the surfaces is the sum of these molecules
acting independently of each other. However it is unlike the chemical synthesis
approach that requires different states of isomerization. The principles
that it relies on are robust and just require
photodissociation of polymer to a binding site, and, as in the case of biological
motors, some degree of asymmetry.
This asymmetry, plus a disruption of thermal equilibrium due to energy input
are the two factors needed to convert energy from chemical or electromagnetic energy, to
mechanical~\cite{ProstPRL}.

The paper is organized as follows. In Sec. \ref{sec:TheSystem} the
photo-mechanical system is described and rough estimates of its
operation are given in Sec. \ref{sec:EstSysPars} including a discussion
of its potential efficiency. In order to illustrate the characteristics
that need to be understood, a three dimensional simulation of this device
is carried out In Sec. \ref{sec:3dModel}. At this point the physics of this system is 
investigated in more detail, starting with general considerations, in Sec.
\ref{sec:SteadyState} of its steady state behavior using Fokker Planck equations.
The results will be used in subsequent sections. An exact description of this
system is possible in steady state and the resulting differential equation
is solved numerically in one dimension in Sec. \ref{sec:1dSolns}. This allows
to understand better how effective the asymmetry in the force produces power,
and influences the direction that is taken in the rest of this work. Sec.
\ref{sec:UnboundEquilMod} studies a useful limit that is hard to probe
numerically, but allows to understand how the efficiency of this system is
related to spring stiffness, metastability, relaxation times, and the
photo-dissociation rate. We then use this model to understand the efficiency,
Sec.  \ref{subsec:EffLargePowerStroke}
using a one dimensional Monte Carlo model that is in this regime. Finally in
Sec. \ref{sec:Conclusions} we conclude on how the physics that has been learned 
might be useful in optimizing experimental parameters for such a system.

\section{The System}
\label{sec:TheSystem}

\begin{figure*}[htp]
\begin{center}
(a) \includegraphics[width=0.4\hsize]{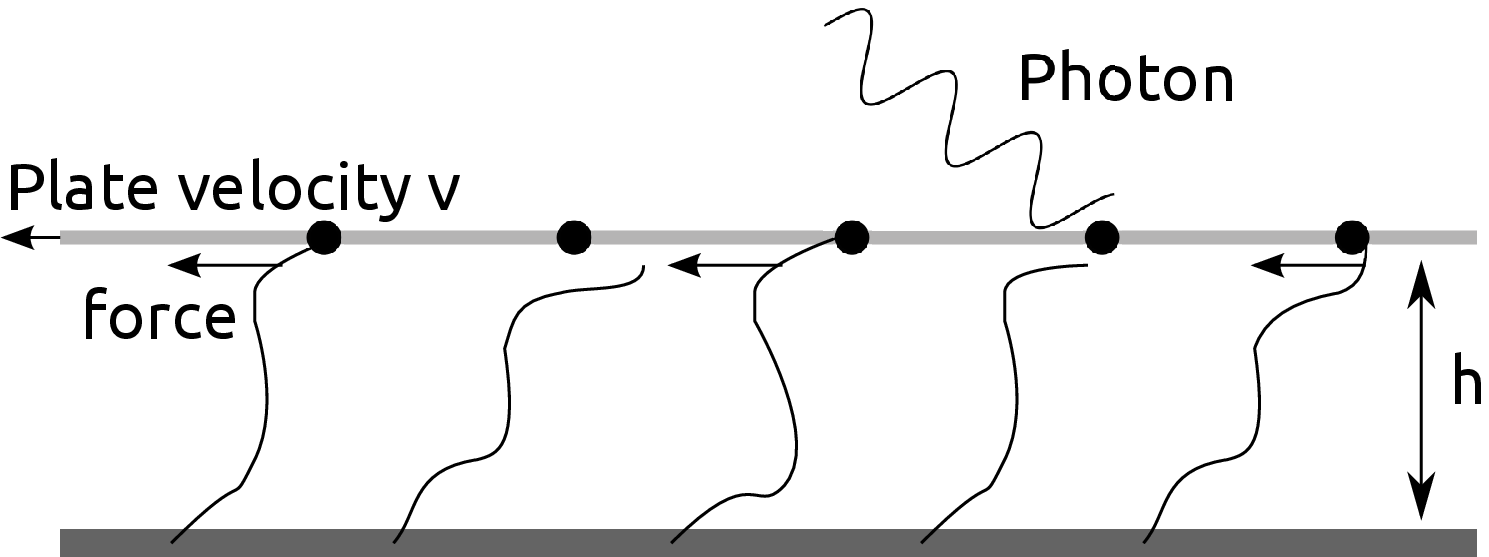}
(b) \includegraphics[width=0.4\hsize]{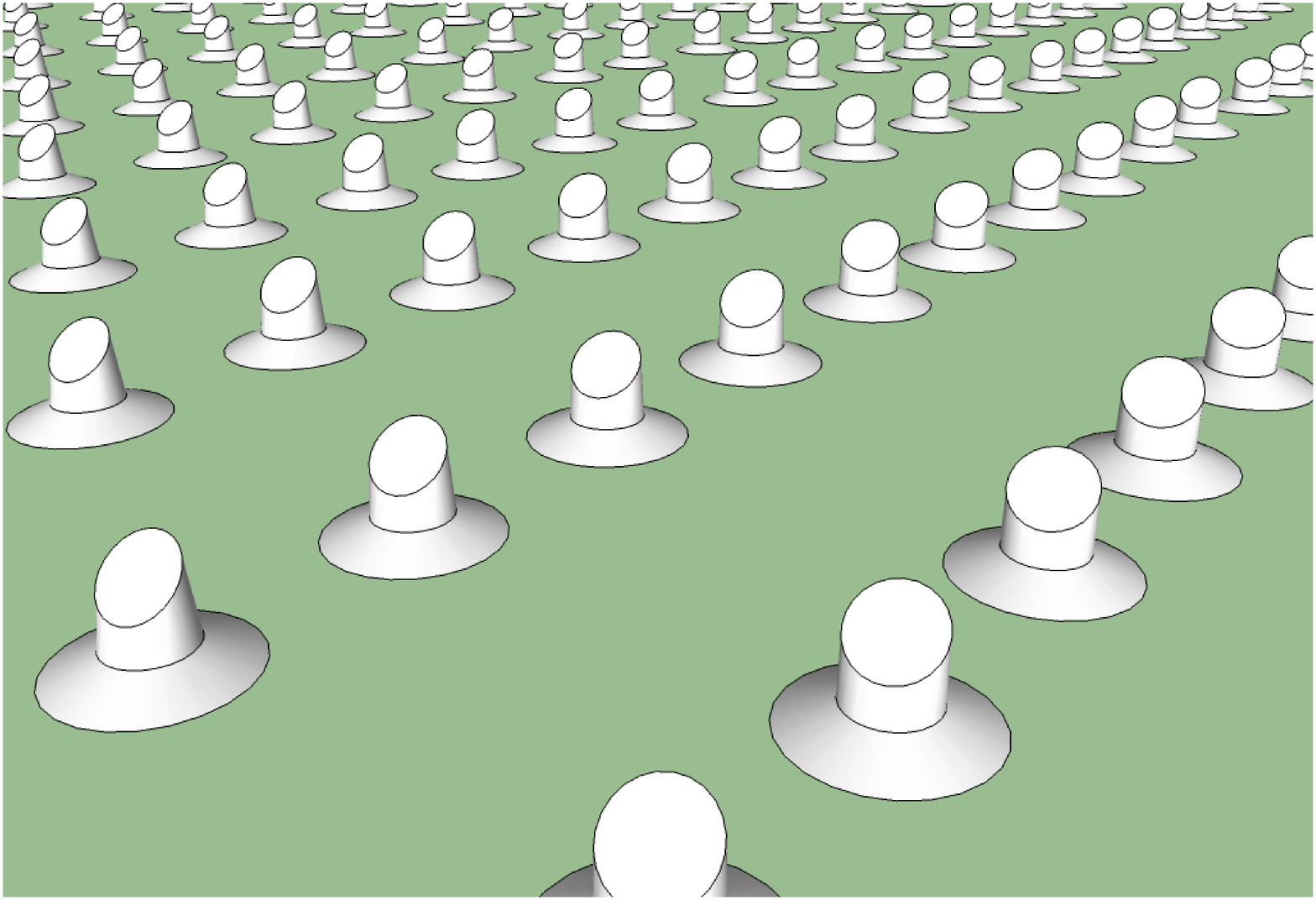}
\caption{
(a) A schematic of the light energy to mechanical converter. A polymer
brush of semiflexible chains is put in contact
with a  plate a height $h$ above the base, containing an array of anisotropic binding sites. Polymer ends are
trapped by unlikely thermal fluctuations, applying a non-zero average force parallel to
the plate. Photodissociation of the polymer ends 
with the binding sites releases the chain end that will move until it finds
another site. The upper plate moves at a velocity $\rm v$, thereby generating
power. A single track is shown, and the separation between binding sites is $L$.
(b) A three dimensional sketch of the binding sites. The binding of the polymer ends
(not shown) takes place on
the oval surfaces protruding from the plate, and do so in an asymmetric manner.
The sites are organized into one dimensional tracks, where the distance between
tracks need not be the same as $L$ in (a).
}
\label{fig:device}
\end{center}
\end{figure*}

The components needed to do constant photomechanical energy conversion are
are illustrated in Fig.  \ref{fig:device}.
\begin{itemize}
\item {\bf A flat plate of semi-flexible polymer brushes}. Polymer brushes are
polymers, each with one end tethered to a wall in a suitable solvent.
\item  {\bf A parallel plate right above the brush}. 
These polymers are put in contact with a parallel
plate close to the surface so that their ends are able to interact with it.
\item {\bf The parallel plate contains an array of photoreactive binding
sites.} It is crucial that these
polymers bind with the surface in an asymmetric way, 
so that the average binding orientation of each is the same, and not perpendicular to the plate.
\item {\bf At least one of the plates is transparent.} Light 
causes the unbinding of polymer ends from the photoreactive binding sites. 
\item {\bf Binding catalyst}. To control the rate at which binding occurs, the
binding of the end of the polymer to a binding site can be facilitated by the
use of a catalyst. The concentration of the catalyst can be used to control the
rate of binding. 
\end{itemize}

In most of the discussion below, the polymers can be considered to be separated
from each other by a sufficient distance so that we can ignore inter-chain interactions.
The absorption of light causing unbinding will happen asynchronously. The ends
tethered to the lower plates are at positions that are either random or
incommensurate with the binding sites of the upper plate. Altogether this means that
the motion of each polymer is uncorrelated with the others in the system.

The total effect of the forces acting on the plates can be used to perform
work against a force acting on the upper plate. Because we assume that
the number of polymers contributing is very large and are asynchronous, the net velocity of the
upper plate $v$ will be constant, as will be the net force acting on the plate.

In the following we will analyze the system and model it in different ways to 
better understand the power generation.

\section{Estimate of system parameters}
\label{sec:EstSysPars}

We will now make a crude estimate with realistic parameters, how well this device should work.
We envisage separate one dimensional tracks of the kind shown in Fig.
\ref{fig:device}(b). The spacing between the tracks could be greater than the
distances between polymers on a single track.  In reality, it might prove
more efficacious to manufacture a square array of polymers and binding sites
but we will allow two separate length scales in our analysis below.

We will assume a solar intensity of $I = 600 W/m^2$, and an average photon energy of  $e = 2eV$. We
will take the relaxation
time of a polymer to be $\tau = 10^{-7}s$, which corresponds roughly to a polymer size of $3nm$. 
The number of unbinding events, if every photon was
absorbed at a binding site, is $I/e \approx 2\times 10^{21}/(m^2 s)$. If the density of polymers is
$\sigma$, then for all of these events to be utilized requires $I/e = \sigma/\tau$, or
$\sigma = 2\times 10^{14}/m^2$, which is a separation of $1/\sqrt{\sigma} \approx 70 nm$.
The amount of energy per step that is gained by a photo-dissociation event is of
order $k_B T$. With such parameters, this is expected to yield an efficiency of
order $1\%$. However, as we will show below, longer relaxation times allow 
for more energy per step, and this requires a higher polymer density. However
the two dimensional density is limited by inter-chain interactions. In the case
considered here, this density could be increased roughly by three orders of
magnitude. However the
relaxation time for this system depends exponentially on the force generated, so 
a three order of magnitude increase in relaxation time
would only increase the efficiency by a factor of about $7$ furthermore, the
detailed analysis presented below suggests that with optimal conditions, the
efficiency of conversion is about $7\%$.
To increase the efficiency of solar conversion further might require stacking devices.

There are other means of increasing the solar conversion efficiency by first converting solar radiation to 
lower energy photons. For example, two methods for doing this are fluorescent down-conversion of solar
photons~\cite{Klampfatis}, or thermophotovoltaic down-conversion~\cite{Harder}.

Further speculation on device efficiency is premature as there will undoubtedly
be many unforeseen technical problems that will likely provide other obstacles
to increasing device efficiency. However an efficiency for direct
photo-mechanical conversion of about $7\%$ is still a useful amount of power
comparable to photovoltaic conversion with amorphous solar cells, and also because
power is lost in electro-mechanical conversion, which is not a problem with
direct mechanical energy conversion.

\section{Three dimensional model}
\label{sec:3dModel}

We start by simulating a three dimensional model of this system. There are two
components to the system, the polymer and the binding sites. The polymer chain
is modeled as having $N$ links of fixed length, and is
semiflexible with chain stiffness $K$. Denote the coordinates of the $ith$ bead as $\br_i$. The elastic potential for the middle of the chain
for the $ith$ bead is
\begin{equation}
U_E(i) \equiv -\frac{C}{2}(|\br_{i+2}-\br_i|^2 + |\br_{i-2}-\br_i|^2)
\end{equation}
where $C$ is the elastic constant.

The binding potential of the polymer has two components, an isotropic component
$U_i$ and directional component $U_d$. $U_i(r)$ is short range with a length
scale $r_s$ and scale $V_a$
\begin{equation}
U_i(r) \equiv -\frac{V_a}{8}(r_s^2-r^2)^4
\end{equation}
and $U_d$ uses a direction $\nhat_a$ so that the difference
between the last two end beads $\delr = \br_1- \br_0$ will give a minimum in $U_d$
along that direction
\begin{equation}
U_d(\br_0, \br_1) \equiv (\bigl|\frac{\delr-\nhat_a}{|\delr-\nhat_a|}\bigr|^2 + 1) U_i(r) .
\end{equation}

The end attached to the lower surface is always bound. The other end can bind to
a periodic linear array of binding sites equally spaced at a distance $L$.

There are two states that the system can be in, unbound, $0$, and bound $1$. 
There are two parameters that control binding, the rate at which dissociation
occurs $c_0$, and the rate that dissociated ends can be rebound $c_1$.

The distance between the lower and upper surface is $h$ and their relative
velocity is $\rm v$.

The model was simulated with a Langevin equation, at finite temperature $T$.
Although inertial effects are typically small at these microscopic scales, it
was included for completeness. An algorithm was used that efficiently updates
this system with fixed link lengths~\cite{DeutschCerfFriction}.

\begin{figure*}[htp]
\begin{center}
(a)
\includegraphics[width=0.4\hsize]{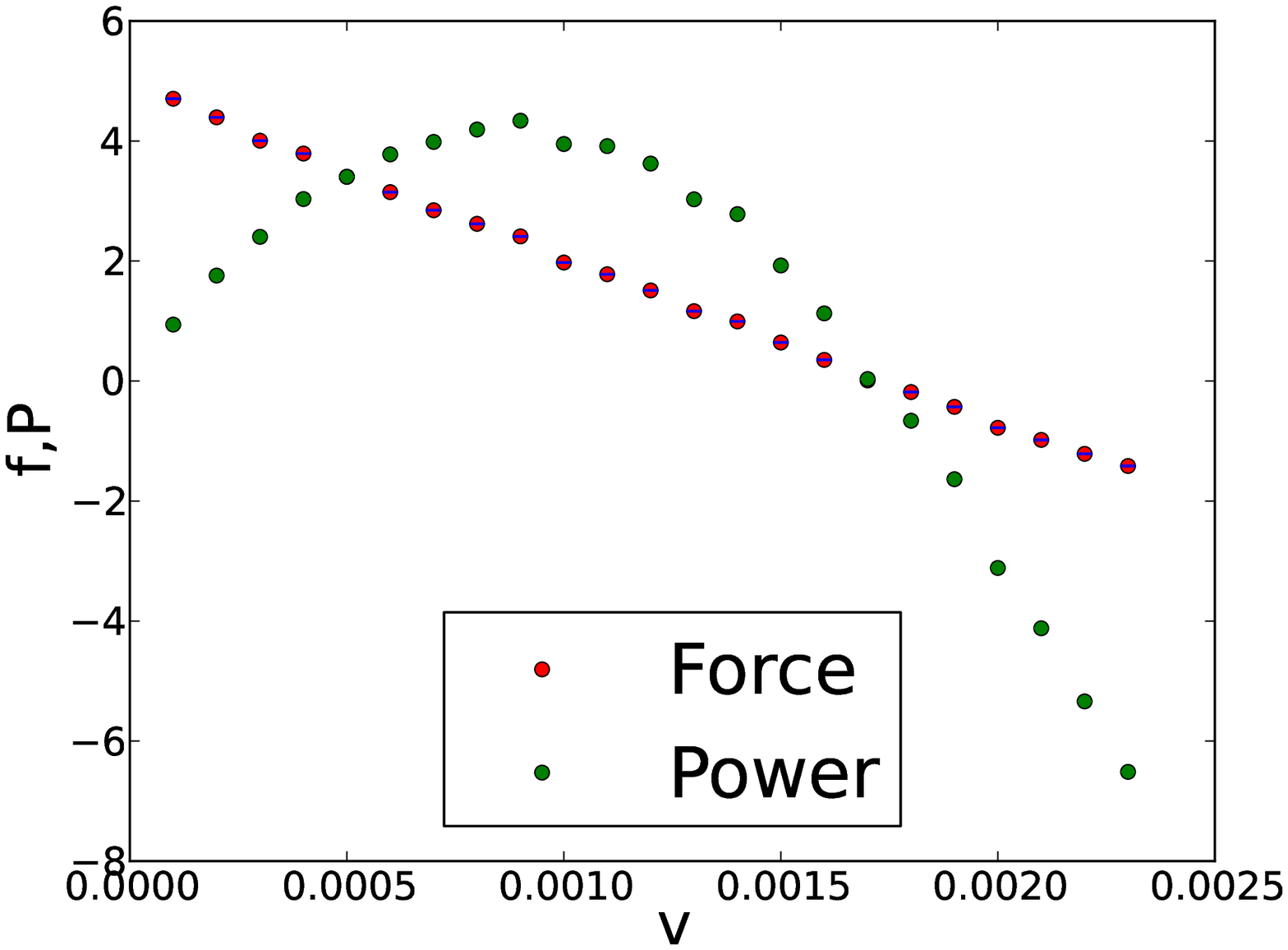}
(b)
\includegraphics[width=0.4\hsize]{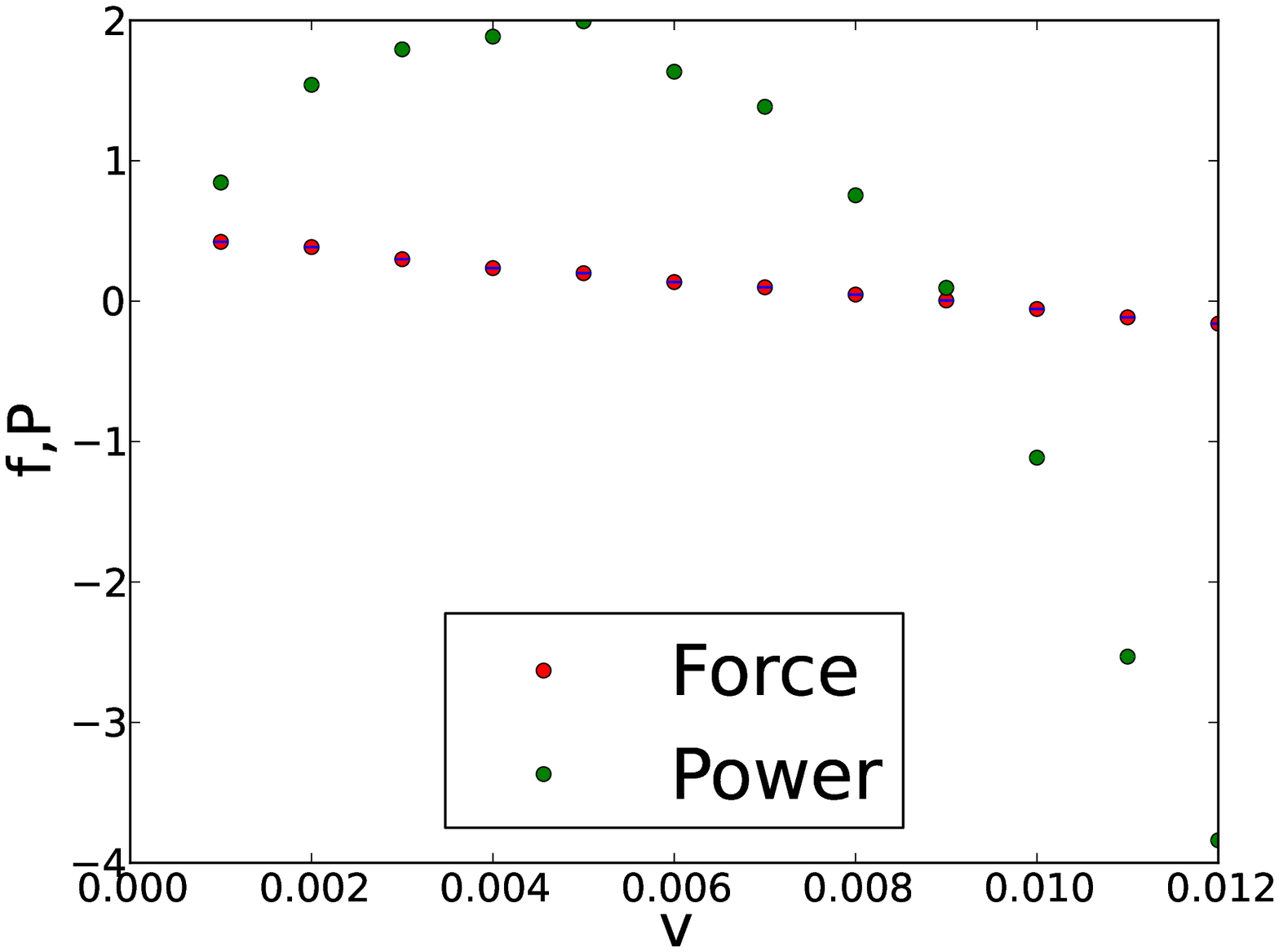}\\
(c)
\includegraphics[width=0.4\hsize]{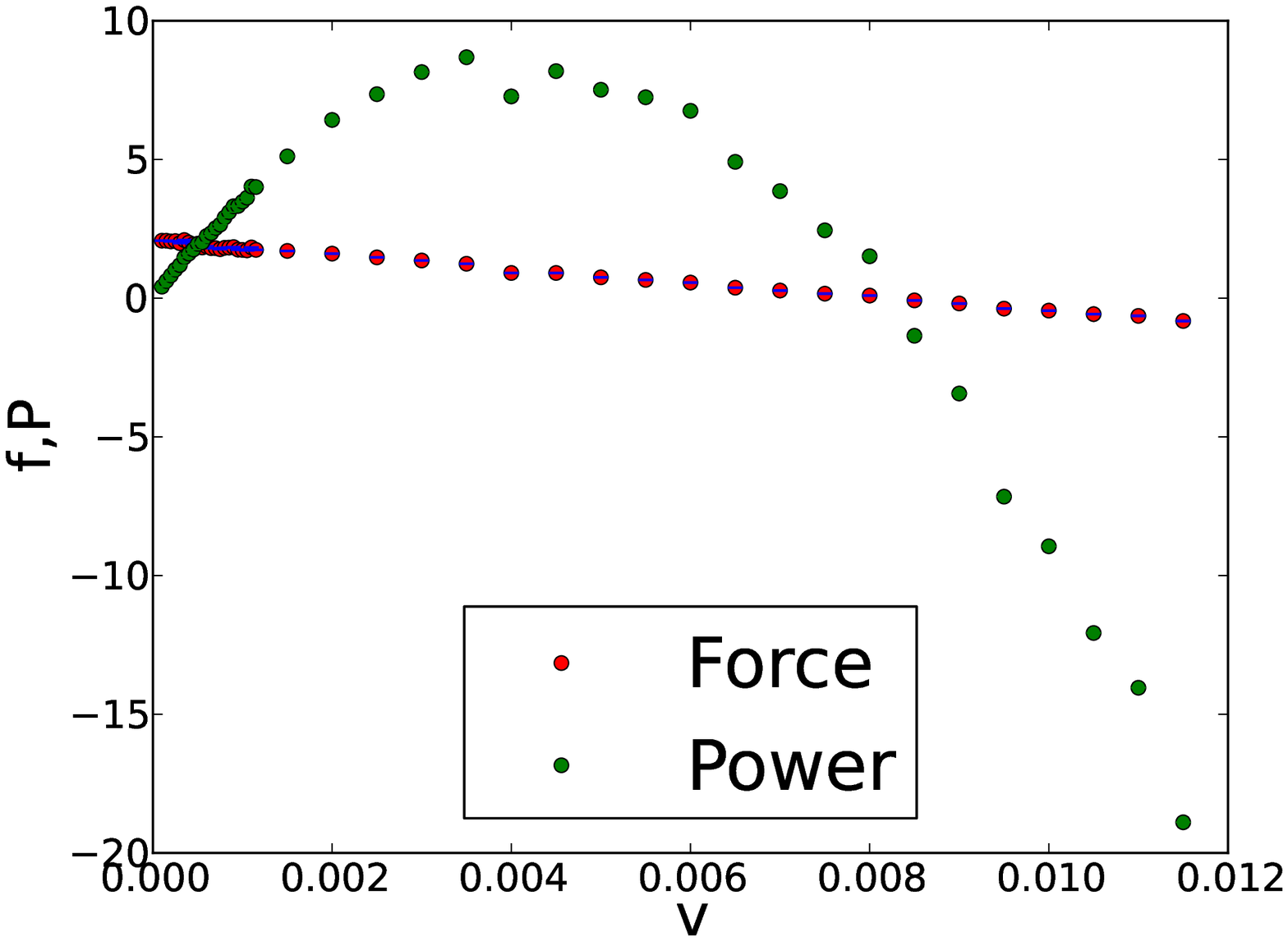}
(d)
\includegraphics[width=0.4\hsize]{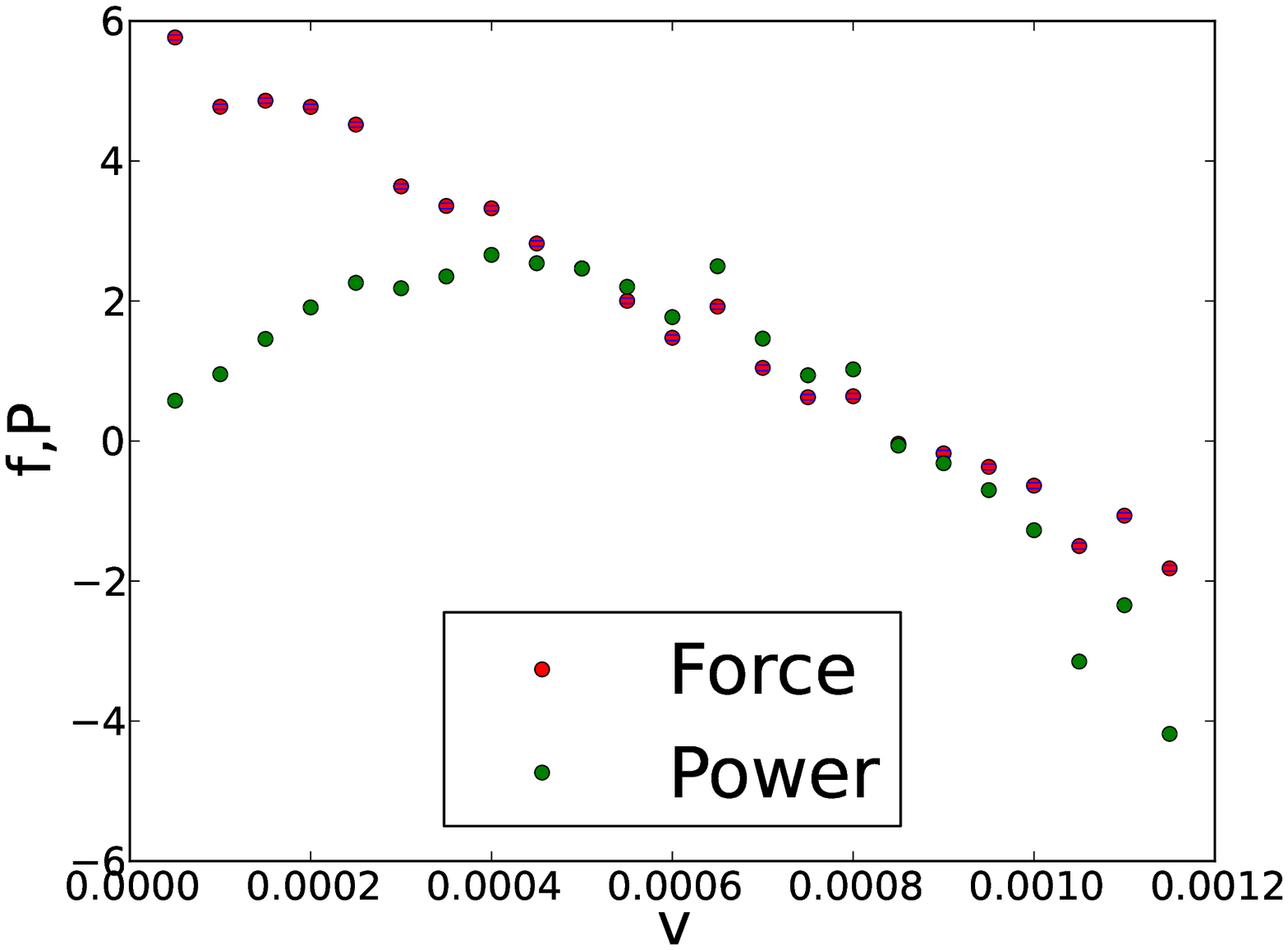}
\caption{
The average force $f$ and the power $P=f {\rm v}$ measured as a function of the
relative velocity between the plates $\rm v$. For clarity, the power is multiplied by $2000$. 
There were $4$ links, with a plate
separation of $h=1$, spacing between binding sites $L = 2$, the direction of $\nhat_a$, that is, the binding angle is
$\pi/4$, $V_a=30$, link length of $1$, and elastic constant $C=1.5$,
potential range $r_s = 2.0$, and a coefficient of damping of $10$. (a) the rate of dissociation
$c_0 = 0.01$, rate of rebinding $c_1 = 0.05$. (b) $c_0 = 0.05$, $c_1 = 0.01$
(c) $c_0 = 0.05$, $c_1 = 0.05$
(d) $c_0 = 0.005$, $c_1 = 0.05$
}
\label{fig:3dfvsv}
\end{center}
\end{figure*}

The simulation was run to determine how the force and power 
generated are influenced by the plate velocity. Fig. \ref{fig:3dfvsv} shows the
results of four simulations with different values of $c_0$ and $c_1$.
The parameters used were: $4$ links, $h=1$, spacing between binding sites $L = 2$, the direction of $\nhat_a$, 
is $\pi/4$, $V_a=30$, link length of $1$, particle mass of $1$, $C=1.5$,
$r_s = 2.0$, and a coefficient of damping of $10$.

The force generated is highest approaching zero velocity and decreases until it
becomes negative, at which point it is taking mechanical energy to move it at
such high speeds. Qualitatively, this is the point where frictional drag
dominates over photo-energy conversion.

Note that the highest power $P=f{\rm v}$ is seen for $c_0 = c_1 = 0.05$. The
force extrapolated to ${\rm v} = 0$ is $f=2.07$ and is about three times
less than the force seen in Fig. \ref{fig:3dfvsv}(b) where $c_0 = 0.05$ and  $c_0 = 0.01$.

Higher maxima in the power are generally seen with increasing $c_0$ and $c_1$.
As an example, the simulation was run for small elastic constant $C=0.5$, at
different values of $c_0$ and  $c_1$, but with all other parameters the same as
above. This higher power is still at the expense of efficiency, as higher
dissociation rates imply a higher photon flux. 

\begin{figure*}[htp]
\begin{center}
(a)
\includegraphics[width=0.4\hsize]{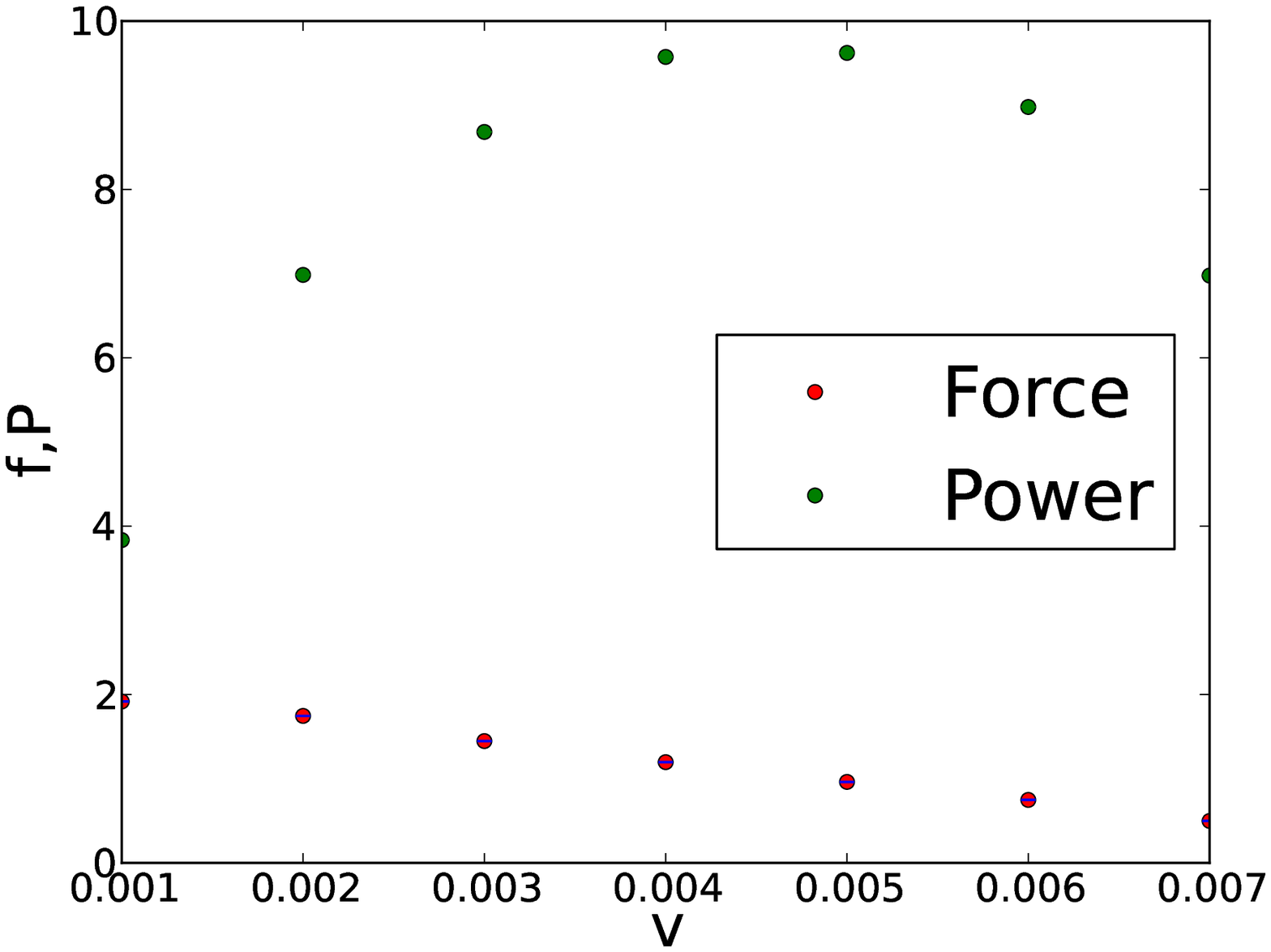}
(b)
\includegraphics[width=0.4\hsize]{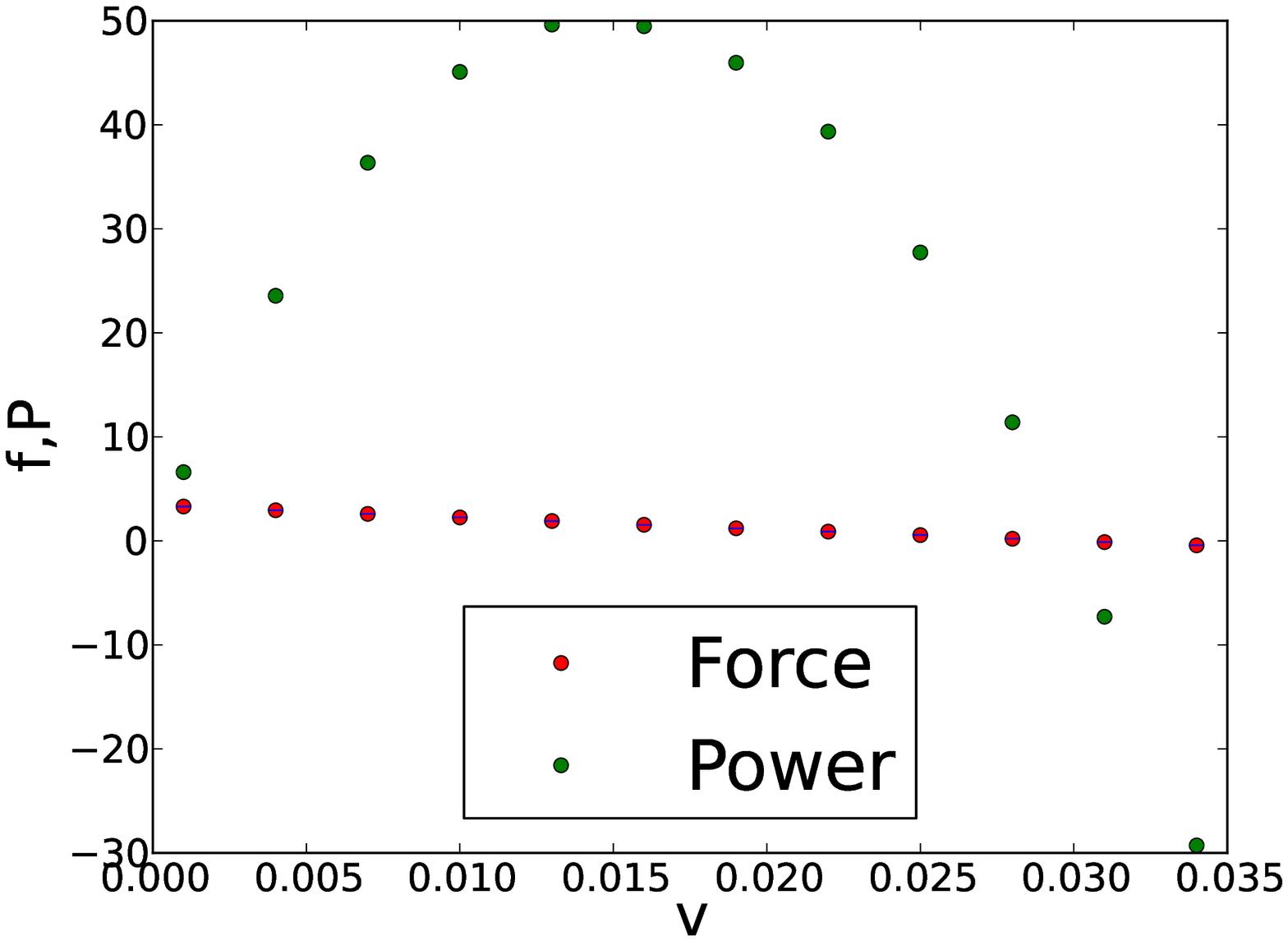}
\caption{
The average force $f$ and the power $P=f {\rm v}$ measured as a function of the
relative velocity between the plates $\rm v$. The same parameters and scaling are used
as in Fig. \ref{fig:3dfvsv} except $C=0.5$ and (a) $c_0=c_1= 0.05$, and (b) $c_0=c_1= 0.4$.
Higher maximum power is seen for larger unbinding rates.
}
\label{fig:3dfvsv2}
\end{center}
\end{figure*}

\section{Steady State Probability Distribution}
\label{sec:SteadyState}

To understand the above simulation in more detail, requires a
better understanding of the mechanisms involved. There have been a very large
number of works on the theory of motor proteins~\cite{ReimannRev,BustamanteKellerOster}, and we will follow the two
state approach mentioned above of Prost {\em et. al.}~\cite{ProstPRL,JulicherRevModPhys}. 
The major difference is that they considered the two states to be periodic
potentials, whereas here we model the system to more closely mimic the
particular device we are investigating allowing the exploration
of force versus velocity. Hence instead of considering the
position of a particle moving between two periodic potentials, we consider
the unbound state to have a free end described as moving in a bound
potential around the tether point. Likewise, because of this tethering, the
bound state potential is not periodic but has a periodic component as we
will describe in detail below.
Here we consider the probability distribution for this system in steady
state, which is described by a Fokker Planck equation.

Initially to calculate the power that is produced, we will concentrate on the low velocity
limit, with the two plates moving so slowly that this motion does not affect
chain conformations appreciably. In this way, we can consider the average force
exerted between the plates by a polymer in steady state in the limit  ${\rm v}=0$ so that the
tether point is not moving. It is most convenient to let the point at which
the polymer is tethered to the bottom plate be a variable parameter $\br' = x' {\xhat}$.
With this point fixed, we consider the distribution of the other end of the
polymer $\br$.  We will assume that the internal dynamics of the chain chain
are much faster than the binding and unbinding rates, so that the only degrees
of freedom are $\br$, and if the polymer is bound, $i=0$ is unbound and $i=1$ is bound. Therefore the
probability distribution of the system can be described by a function
$P_s(\br;\br')$. The equations describing this are~\cite{ProstPRL}
\begin{subequations}
\label{eq:FP}
\begin{align}
\partial_t P_0(\br) = \nabla \cdot (\nabla - \bof_0)P_0(\br) - c_0 P_0(\br) + c_1 P_1(\br)\label{eq:FP0}\\
\partial_t P_1(\br) = \nabla \cdot (\nabla - \bof_1)P_1(\br) + c_0 P_0(\br) - c_1 P_1(\br)\label{eq:FP1}
\end{align}
\end{subequations}
where the last $\br'$ argument of $P$ has been left out for notational simplicity.
The units here absorb the diffusion coefficient $D$ together with the time, that
is $t$ as used here and below is really $D$ times the time. The force terms have 
absorbed a temperature factor $T$, that is $\bof_i$, is really the
force times $1/k_B T$.
Those forces, $\bof_0$ and $\bof_1$ are the total forces acting on the upper end
of the chain. When the system it is unbound is the force of a (possibly) nonlinear
spring 
\begin{equation}
\bof_0 = \bof_s(\br-\br'). 
\end{equation}
When the system is bound we have total force
is the sum of the spring force and a periodic force representing the binding
potential 
\begin{equation}
\bof_1= \bof_s(\br-\br') + \bof_p(\br).
\end{equation}
These forces are assumed to be conservative, $\bof_s$ and $\bof_p$ are
derived from potentials that respectively are $V_s$ and $V_p$.

In steady state the left hand sides of these equations are zero. We can
eliminate the last two terms on the right hand side to obtain.
\begin{equation}
\label{eq:ohat1ohat2}
\Ohat_0 P_0 + \Ohat_1 P_1 = 0
\end{equation}
where 
\begin{equation}
\label{eq:Ohatdefn}
\Ohat_i  = \nabla \cdot (\nabla - \bof_i), ~~~ i=0,1.
\end{equation}
We can now calculate the average force exerted by the two potentials, is zero. In steady
state, we multiply Eq. \ref{eq:ohat1ohat2} by $x$ and integrate with respect to $x$, $y$
and $z$. Then using integration by parts, we have boundary terms at infinity.
Because we are assuming that the spring potential grows without bounds, this
confines $P_0$, and $P_1$, to a neighborhood around $\br'$, so that the boundary
terms vanish. We are then left with
\begin{equation}
\label{eq:P0plusP1}
\langle \bof\rangle = \int (\bof_0 P_0 + \bof_1 P_1) d^d\br = 0
\end{equation}
as is expected because $r$ is confined to a region of space so that the average
velocity, and hence average force, will be zero. 

We can also determine the total fraction of time spent in the bound or unbound
states in steady state. First, because the probability of being in any state is unity,
\begin{equation}
\int (P_0(\br) + P_1(\br)) d^d\br = 1 .
\end{equation}
Then by integrating Eq. \ref{eq:FP0} over all space, the derivative term
integrates to 0, giving
\begin{equation}
\int (-c_0 P_0(\br) + c_1 P_1(\br)) d^d\br = 0
\end{equation}
hence
\begin{equation}
\label{eq:norms}
\int P_0(\br) d^d\br = \frac{c_1}{c_0+c_1}, ~~~
\int P_1(\br) d^d\br = \frac{c_0}{c_0+c_1} .
\end{equation}

To obtain the power produced by this device, we are not interested in the total
force acting on the upper chain end because this includes the binding potential, but the
average force due to the spring acting on the lower plate $\langle f_s\rangle$. We would like
to calculate the work done in moving the lower point $\br'$ by one period of
$\bof_p$. After moving one period the system is statistically identical to its
starting point, and this method can therefore give the work done in moving $n$
such periods. Denoting the period of $\bof_p$ by $L$, we would like to calculate
\begin{align}
\label{eq:defnWL}
W_l &= \int_0^L \xhat\cdot \langle \bof_s\rangle dx' \nonumber\\
      &= \int_0^L \int \xhat\cdot \bof_s(\br,x'\xhat)(P_0(\br,x'\xhat)+P_1(\br,x'\xhat))d^3\br dx'
\end{align}
and using Eq. \ref{eq:P0plusP1} 
\begin{equation}
\label{eq:WL}
W_l = -\int_0^L \int \xhat\cdot \bof_p(\br)P_1(\br,x'\xhat)d^3\br dx' .
\end{equation}

In thermal equilibrium, where the transition rates $c_0$ and $c_1$ are both zero, 
we recover the Gibbs distribution. Let us assume, that the system starts, and
therefore remains in state $i=1$. Then
\begin{equation}
P_0(\br, \br') d^3\br = \frac{e^{-V_1}}{Z} d^3\br
\end{equation}
where the partition function 
\begin{equation}
Z(\br') = \int e^{-V_1} d^3\br 
\end{equation}
which will also have periodicity of $L$.
In this case $W_L$ can be easily calculated because $P_0 =0$ and $f_s = -\nabla
V_s(\br-\br') = \nabla'V_s(\br-\br')$. So 
\begin{eqnarray}
\label{eq:Weq0}
W_L &=& \int_0^L \int (\partial_{x'}V_s(\br,x'\xhat)) \frac{e^{-(V_s(\br-\xhat x')+V_p(\br))}}{Z(x')} d^3\br dx'\nonumber\\
    &=& -T (\log Z(L) - \log Z(0)) = 0
\end{eqnarray}
as it must be by the second law of thermodynamics.

\section{One dimensional solution}
\label{sec:1dSolns}

In order to investigate how the power conversion depends on the forces acting on
this system, it is important to simplify the three dimensional model to obtain
a minimal model that depends on far fewer parameters. Therefore we investigate
this model in one dimension.

A key point to understand is how asymmetry in the form of the force produces
power. With symmetric spring and binding potentials, it is easily seem by
symmetry, that no net power can be produced from this system. We now ask how
asymmetry affects the results. We will see that even with a large asymmetry
in the spring potential, the work defined by Eq. \ref{eq:defnWL} is very small.
Simulations using Monte Carlo or Langevin equations are too noisy to provide
good estimates. We therefore instead use a more analytical approach.

The coupled Focker Planck Eqs. \ref{eq:FP} (a) and (b) can be solved to produce
an equation only involving one distribution function, $P_0$ in steady state.
Using the definitions in Eq. \ref{eq:Ohatdefn}, we can eliminate $P_1$.
\begin{equation}
(\Ohat_1\Ohat_0 - c_1 \Ohat_0 - c_0 \Ohat_1) P_0 = 0
\end{equation}
which is a fourth order equation in spatial variables.

Now we restrict the analysis to one dimension. In this case, $\Ohat_i = \px(\px-f_i)$
so we can integrate with respect to $x$. We note that because the spring
confines $P_0$ to a localized region, it will go to zero as
$x\rightarrow\pm\infty$. Therefore the integration constant must also be zero
\begin{equation}
\label{eq:1dDiffEq}
((\px-f_1)\px(\px-f_0) - c_1 (\px-f_0) - c_0 (\px-f_1)) P_0 = 0
\end{equation}
which is a third order linear differential equation.

Eq. \ref{eq:1dDiffEq} was solved by the shooting method~\cite{ShootingMethod}.
The boundary conditions were obtained by considering the system far from $x'$ where $P_0$ is very small. In that
domain, $f_p$ was artificially cutoff so that $f_0=f_1=f_s$. Because the
potential there is no longer changing between the two states, the solution is
that of a system in thermal equilibrium. The solutions were required to match to
these thermal solutions in this regime, far from $x'$.
The equation was solved with three different initial conditions. An appropriate
linear combination of these were constructed to match the boundary conditions as
just described.

The periodic binding potential that was used is
\begin{equation}
\label{eq:Vp}
V_p = A_1 \cos(x) - A_2 \sin(2 x) ~~.
\end{equation}

\begin{figure}[htp]
\begin{center}
\includegraphics[width=\hsize]{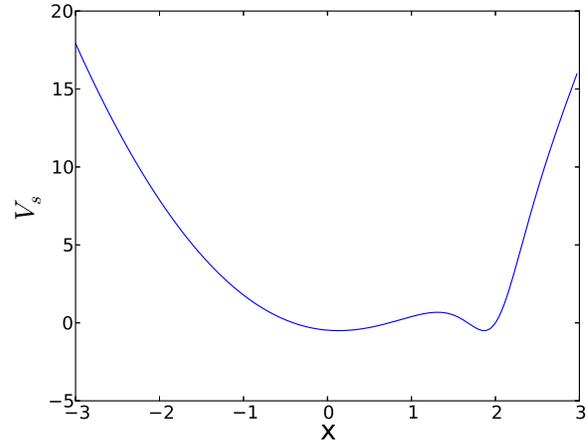}
\caption{
An asymmetric spring potential used to tether the chain. It has a second dip at
approximately $x=2$
}
\label{fig:vs}
\end{center}
\end{figure}

We first consider asymmetry in $V_s$ but with symmetric functions $V_p$, that
is, $A_2 =0$ in Eq. \ref{eq:Vp}.

\begin{equation}
\label{eq:nlfs}
V_s(x) \equiv \half k x^2 - \frac{a}{1+(b (x-d))^2} .
\end{equation}
The first term describes a linear spring with spring constant $k$, the second
adds an asymmetric dip. The parameters are chosen so that this dip is close
in potential to the one created predominantly by the linear term, $k=4$, $a=8$,
$b = 2$, and $d=2$. The function is plotted in Fig. \ref{fig:vs}.

\begin{figure}[htp]
\begin{center}
\includegraphics[width=\hsize]{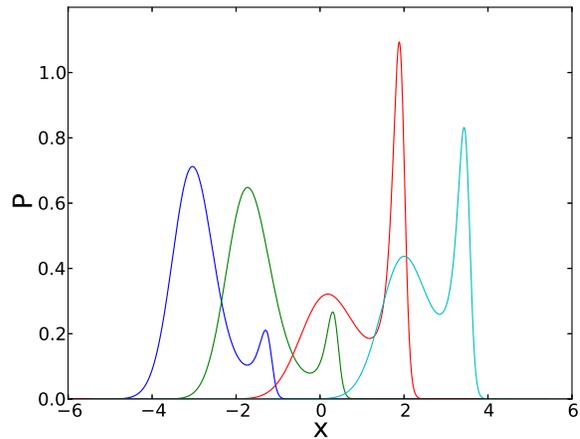}
\caption{
Plots of the probability distribution of $P_1$ as a function of position $x$
for different values of $x'$ for the asymmetric spring model Eq. \ref{eq:nlfs}.
}
\label{fig:Pnlspr}
\end{center}
\end{figure}

Plots are shown in Fig. \ref{fig:Pnlspr} of $P_1(x)$ for four values of $x'$ within a period.

By integrating using these distributions, Eq. \ref{eq:WL}, the work can be
obtained.  With $c_0 = 0.025$ and $c_1 = 0.05$ the work $W_L =   0.00080954$.
With the $c_i$'s $10$ times those values, 
$c_0 = 0.25$, and $c_1 = 0.5$, $W_L =  0.0025$. 

\begin{figure}[htp]
\begin{center}
\includegraphics[width=\hsize]{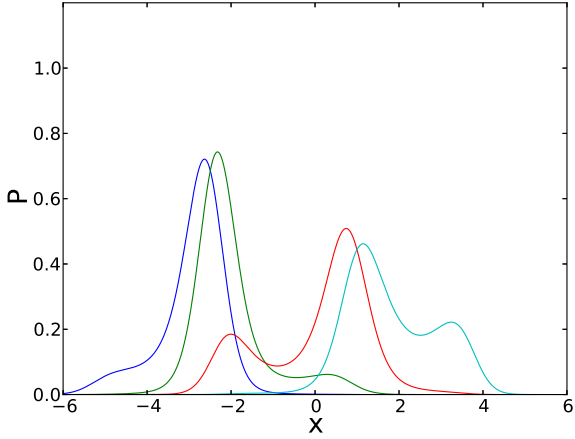}
\caption{
Plots of the probability distribution of $P_1$ as a function of position $x$
for different values of $x'$ for an asymmetric periodic potential, but a linear
spring.
}
\label{fig:Pasvp}
\end{center}
\end{figure}

Now we consider the case of a linear spring so that the nonlinear parameter $a=0$
in Eq. \ref{eq:nlfs}. Instead we make the periodic potential asymmetric by
setting  $A_1 = A_2 =2$ in Eq. \ref{eq:Vp}, still with
$c_0 = 0.25$, and $c_1 = 0.5$. 
Fig. \ref{fig:Pasvp} plots $P_1(x)$ for four values of $x'$ within a period.
In this case, the work $W_L = -0.1596$.

What the numerical results have shown is that asymmetry in the spring potential
is quite ineffective at producing work, whereas asymmetry in the periodic
binding potential is much more effective. We shall use this result below and concentrate
on systems with symmetric spring potentials, but asymmetric potentials.

\section{Unbound Equilibration Model}
\label{sec:UnboundEquilMod}

It is worthwhile to
understand in more detail, what constrains the maximum force of photo-mechanical conversion by
tuning the potentials employed and the binding rate. And to ask how the design will depend on 
the flux of photons. We are limited in our choice of the potentials $V_0$ and
$V_1$ that both must be bounded. In this system, the unbound potential is not periodic which
limits the amount of power that can be generated. In cases considered earlier~\cite{ProstPRL}, where both
potentials are periodic, it is possible to get much more efficient motion by
alternating between states with different potential maxima. However this is not
relevant to our system.

To understand this better, it is useful to examine a limit where we can treat
the system analytically. We therefore examining the case where the 
binding rate of unbound chain is sufficiently that we can regard it in thermal
equilibrium. 

\begin{figure}[htp]
\begin{center}
\includegraphics[width=\hsize]{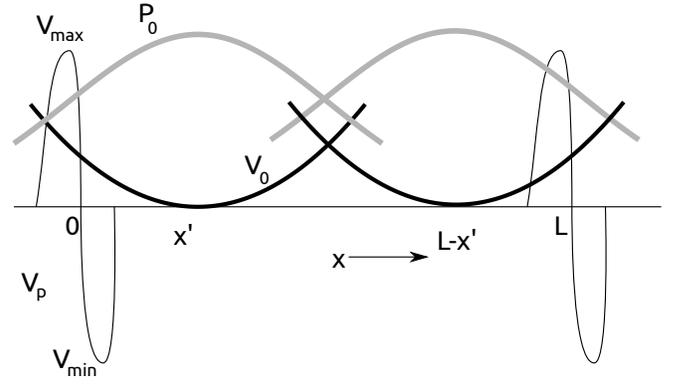}
\caption{
Illustration of the kind of potentials employed in the Unbound Equilibration
Model. The periodic potential $V_p$ is nonzero only on small regions on the
x-axis. It shows a large peak of height $V_{max}$ and large negative dip
$V_{min}$. There spring potential $V_0$ is parabolic, except that it has a
cutoff where it becomes infinite when stretched by more than $l_c$. The
corresponding probability distribution $P_0$ (grey curve) is assumed to have relaxed to
equilibrium. The spring potential is shown at two different position, when it
is centered at $x'$ and $L-x'$.
}
\label{fig:1dpot}
\end{center}
\end{figure}

The model is illustrated in Fig. \ref{fig:1dpot}. We take the unbound potential
to be that of a linear spring below a cutoff
\begin{equation}
\label{eq:v0cutoff}
V_s = V_0 =\begin{cases}
k x^2/2 , & \text{if $-l_c < x< l_c$}.\\
\infty, & \text{otherwise}.
\end{cases}
\end{equation}
we choose the spring coefficient $k$ and the cutoff length $l_x$ such that
$l_c \exp(-k l_c^2/2) \ll 1$.  Below, we will take $l_c = L/2$ or $\infty$.

The periodic potential $V_p$ is taken to be localized at periodic points (with
a separation of $L$) that rapidly vary from a large maximum $V_{max}$ to a
minimum $V_{min}$, as shown. We take the region over which this happens to be
negligibly small compared to other length scales in the problem.

To understand how equilibration can occur in this model, we can rewrite 
Eqs. \ref{eq:FP} in the steady state limit
\begin{subequations}
\label{eq:SFP}
\begin{align}
\nabla \cdot (\nabla - \bof_0)P_0(\br) - c_0 P_0(\br) = -c_1 P_1(\br)\label{eq:SFP0}\\
\nabla \cdot (\nabla - \bof_1)P_1(\br) - c_1 P_1(\br) = -c_0 P_0(\br) \label{eq:SFP1}
\end{align}
\end{subequations}
Although this is time independent, we consider the related time dependent equations for
variables $\tilde{P}_i(\br, t)$
\begin{subequations}
\label{eq:LFP}
\begin{align}
 \partial_t \tilde{P}_0(\br)-\nabla \cdot (\nabla - \bof_0)\tilde{P}_0(\br)  = 0 \label{eq:LFP0}\\
 \partial_t  \tilde{P}_1(\br)-\nabla \cdot (\nabla - \bof_1)\tilde{P}_1(\br)  = 0 \label{eq:LFP1}
\end{align}
\end{subequations}
where $P_i$, $i=0,1$, is the Laplace transform of $\tilde{P}_i$, $P_i(s) = {\cal L} \{\tilde{P}_i\}$, where the conjugate variable
for equations (a) and (b) are $s =c_0$ and $s = c_1$, respectively. The initial conditions on each equation are 
\begin{subequations}
\label{eq:IFP}
\begin{align}
\tilde{P}_0(\br,t=0) = c_1 P_1(\br,s=c_1) \label{eq:IFP0}\\
\tilde{P}_1(\br,t=0) = c_0 P_0(\br,s=c_0) \label{eq:IFP1}
\end{align}
\end{subequations}

These equation have a direct physical interpretation. The left hand sides in
Eqs. \ref{eq:LFP} are those of particles in diffusing in potentials but with
conservation of particles. For long enough times, independent of initial
conditions, the solution to these equations will go to thermal equilibrium given
by the Gibbs distribution. 
Eq. \ref{eq:LFP0} describes diffusion in a quadratic potential. We require that
we are probing this at long enough times $t$, so that it will have nearly
reached this equilibrium state. Call this longest relaxation time $\tau_0$.
The solution to the diffusion equation $\tilde{P}(\br,t)$ for either of Eqs.
\ref{eq:LFP}, can be written as a sum over spatial
eigenfunctions, $\phi_n(\br)$ that decay at different rates $\lambda_n$,
(arranged to be monotonically increasing):
\begin{equation}
\label{eq:phiexpansion1}
\tilde{P}(\br,t) = \sum_{n=0}^\infty \phi_n(\br) \exp(-\lambda_n t), 
\end{equation}
of which the smallest $\lambda$, $\lambda_0 = 0$, corresponds to the equilibrium state. The relaxation
time is $\tau_0 = 1/\lambda_1$, Thus the Laplace
transformed variable $P$ can be written
\begin{equation}
\label{eq:phiexpansion2}
P(\br,s) = \sum_{n=0}^\infty \phi_n(\br) \frac{1}{s+\lambda_n} .
\end{equation}
If the $n=0$ term is to dominate, we therefore require $s = c_0 \ll 1/\tau_0$.
The physical interpretation of this condition is that binding typically occurs only after
many relaxation times of the unbound end.

$P_0(\br, s=c_0)$ will be dominated by the first term in Eq. \ref{eq:phiexpansion2}
and therefore proportional to the eigenfunction $\phi_0(\br)$ which is
the equilibrium distribution and is $\propto \exp(-V_0)$.
An important simplifying point in the above approach is that the precise form of
the initial condition Eq. \ref{eq:IFP0} is not not important, but because of
particle conservation, only the total area under $P_1$ affects the result.

Now we consider the solution for $P_1$. If we consider unbinding times much
longer than the relaxation time in the bound state, we also arrive a thermal
equilibrium, which as was shown by Eq. \ref{eq:Weq0} to lead to no work being performed. Instead
we will consider situations where there are very long lived metastable states.
In Fig. \ref{fig:1dpot}, the potential seen by a particle in Eq. \ref{eq:LFP1}
is $V_1 = V_0 + V_p$. If a particle starts between $0$ and $L$, it will remain
trapped in that region for a Kramer's time which (ignoring algebraic pre-factors)
depends on $x'$, but has a minimum value of $\tau_m \propto \exp(V_{max})$. By
choosing large enough $V_{max}$ this can be made arbitrarily long. 

A particle in such a metastable state will relax to a metastable equilibrium,
obeying Eq. \ref{eq:phiexpansion1} that will eventually fail for times $t >
\tau_m$. In this expansion, the longest relaxation time to this metastable state
$\tau_1 = 1/\lambda_1$,
will be taken to be much smaller than $\tau_m$. Thus the above argument on the
range of $c_0$ can be used {\em mutatis mutandis} to restrict
the unbinding rate to $1/\tau_m \ll c_1 \ll 1/\tau_1.$

In this regime we can understand the solution to  Eqs. \ref{eq:LFP1} and \ref{eq:IFP1} 
by considering the corresponding Greens function $\tilde{G}(\br,t;\br_0)$. We replace the initial
condition Eq. \ref{eq:IFP1} by 
\begin{equation}
\label{eq:GIFP} 
\tilde{G}_1(\br,t=0;\br_0) = \delta(\br-\br_0) . 
\end{equation}

We can then obtain $\tilde{P}_1$ from the Greens function from
\begin{equation}
\label{eq:P1intGP0}
\tilde{P}_1(\br,t) = \int \tilde{G}_1(\br,t;\br_0) c_0 P_0(\br_0, s = c_0) d^d\br_0 .
\end{equation}

For a wide range of times, and regions of $\br'$, the $\tilde{G}_1$ will approach the same
metastable state. A particle starting at any point within a certain region will
end up stuck in the same state and hence approach the same metastable equilibrium. 
Confining our attention to the one dimension model of Fig.
\ref{fig:1dpot}, for $0 < x_0 < L/2$ the solution will be strongly localized at
the minimum $x=0$. For $L/2 < x_0 < L$ the effects of $V_p$ are negligible
and $\tilde{P}_1(x) \propto \exp(-V_0)$. This is because the potential $V_0$ 
in Eq. \ref{eq:v0cutoff} is cutoff which does not allow the particle to visit
the $x=0$ region. Hence the effect of $V_p$ is only seen close to $x=L$ where
it provides a strong repulsion, but over a negligibly small region of $x$.
For $-L/2 < x_0 < 0$, $V_p$ also does not contribute.
For $L < x_0 < 3L/2$ the solution will be strongly localized at
the minimum at $x=L$. Because of particle conservation, the area under
$\tilde{G}_1$ is always unity.

A physical interpretation of the above equations in terms of a one dimensional one particle system 
can now be made using the Laplace transformed
variables and the metastable limit considered above is now apparent.
In the unbound state, the
particle reaches thermal equilibrium relaxing to the Gibbs
distribution, $P_0 \propto \exp(-V_0)$. Then the periodic potential $V_p$ is suddenly
added in. The position at the time of switching is labeled $x_0$. Depending
on which interval $x_0$ is in, $x$ will equilibrate to the corresponding
metastable equilibrium with $P_1 \propto \exp(-V_1)$ and is completely confined to
that interval. The relative probabilities
being in one of the three above regions is obtained by the area under $P_0$ for
that interval.

Now we know how to determine $P_0$ and $P_1$, we would like to calculate the
work $W_L$ given by Eq. \ref{eq:defnWL}. Because of the symmetric form assumed
for $f_s$, the $f_0 P_0$ term in the integrand gives zero contribution and we
are left with
\begin{equation}
\label{eq:symWL1d}
W_l = \int_0^L \int_{-\infty}^\infty  f_0(x-x') P_1(x,x')dx dx'
\end{equation}
and consider first how to calculate the inner integral
\begin{equation}
\label{eq:F1d}
f_l(x') \equiv \int_{-\infty}^\infty f_0(x-x') P_1(x,x')dx 
\end{equation}
where $x'$ is the position of the tethered end.  Eq. \ref{eq:v0cutoff} gives $f_0(x) = -kx$ for $|x|<L/2$.

Using the prescription we have found for $P_1(x,x')$, which is simplified by the
above physical interpretation, we can partition this integral
into the different $x$-intervals of metastability: $I_- \equiv[-L/2,0]$, $I_0 \equiv [0,L]$, and $I_+ \equiv [L,3L/2]$.
The value of $P_1(x,x')$ depends on the probability of initially being trapped in
one of those three intervals. Because of the cutoff we have imposed on $V_0$,
only two intervals need be considered for a given value of $x'$. For $0 < x' < L/2$, only intervals $I_-$ and
$I_0$ occur.  The probability that $x \in I_-$ given $x'$ is
\begin{equation}
E(x') \equiv Prob(x \in I_-| x') = \int_{-\infty}^0 p_0(x-x') dx  = \int_{x'}^\infty p_0(x) dx
\end{equation}
and the probability that $x \in I_0$ that is, $P(x \in I_0|x') = 1-E(x')$.
Here $p_0(x)$ is proportional to $P_0(x)$ but normalized to unity, to simplify
the presentation.
We can obtain the values for $L/2 < x' < L$ by symmetry, so that $P(x \in I_+| x') = E(L-x')$ and
$P(x \in I_0|x') = 1-E(L-x')$.  $E(x)$ is simply related to the complementary
error function in the limit considered here, where the effects of the cutoff in
the potential will have negligible effect, $E(x) = \frac{1}{2}erfc(\sqrt{k}x)$.

There is a symmetry in many of the quantities considered, 
as shown in Fig. \ref{fig:1dpot} where the
potential $V_0$  and corresponding probability distribution $p_0$ is for the tethering point at
$x'$ and at $L-x'$. Therefore it is convenient to consider $f_l(x') + f_l(L-x')$. 
This can be written as
\begin{eqnarray}
\label{eq:flplusfl}
-\frac{k c_0}{c_0+c_1} (-&E&(x')\langle x \rangle' +  E(x') x' \nonumber\\
       - (&1&-E(x'))x' + (1-E(x'))\langle x\rangle'')
\end{eqnarray}
where
\begin{equation}
\langle x \rangle' \equiv -\frac{\int_{-\infty}^0 (x-x')p_0(x-x')dx}{\int_{-\infty}^0 p_0(x-x')dx} 
= \frac{\int_{x'}^\infty x p_0(x)dx}{E(x')}
\end{equation}
and
\begin{equation}
\langle x \rangle'' \equiv -\frac{\int_{-\infty}^0 (x+x')p_0(x+x')dx}{\int_{-\infty}^0 p_0(x+x')dx} 
= \frac{-E(x')}{1-E(x')} \langle x \rangle' .
\end{equation}

The four terms in Eq. \ref{eq:flplusfl} correspond to contributions from respectively regions $I_-$, $I_+$ , $I_0$, and
$I_0$. The first and third term are contributions from $f_l(x')$ and the others
are from $f_l(L-x')$. The factors involving $c$ give the correct normalization
according to Eq. \ref{eq:norms}. Combining the above equations, 
\begin{equation}
W_L = \frac{2 k c_0}{c_0+c_1} \int_0^{L/2} 2 E(x')(\langle x\rangle' - x') + x' dx' .
\end{equation}

In the limit of large $L$, which we are considering by virtue of the condition
on the cutoff in $V_0$ imposed by Eq. \ref{eq:v0cutoff}, the integrand
simplifies because $\langle x\rangle'$ becomes exponentially close to $x'$, and
the only term remaining is $x'$. Thus for large $L$, $W_L = (c_0/(c_0+c_1))k L^2/4$
The factors involving the $c$'s represent the fraction of time spent in the
bound configuration. The last term increases quadratically with $L$. This result
is misleading if not taken with the appropriate limits that have been assumed in
its derivation. The factor $(c_0/(c_0+c_1))$ is very close to unity as we are
assuming that the relaxation time in the unbound state is much faster than in
the bound state, hence $c_0 \gg c_1$. However the work $W_L$ to move a distance
$L$ was assumed to be in the adiabatic limit, and here the time scales
associated with bound state relaxation are exponentially long. This is because
to reach this metastable equilibrium the particle has to hop over barriers of
size $V_0(x')$, see Fig. \ref{fig:1dpot}. Hence the  longest relaxation time for this is
at $x'=L/2$ and is of order $\exp(kL^2/4)$. Note that this does not contradict
our assumption that we are still in a region of metastability, which requires
times much less than $\exp(V_{max})$. But for this to work, we require that $V_{max} \gg V_0(L/2)$

Now we consider the case where the spring length cutoff $l_c \rightarrow \infty$. The
disadvantage of this is that the chain end can, in principle, hop over many
barriers ending up arbitrarily far from the tether point, and that these hopping
times should be included in the above analysis. However, the probability of such
a hop becomes negligible when the probability of finding the chain end there is small. Hence we
still have a clear separation of time scales between metastable states as
discussed above, and fully equilibrated system, which requires surmounting the
energy barrier $V_{max}$. In this case, we can therefore assume that when the
chain end $x'$ is between $nL$ and $(n+1)L$, it will strongly localized at $x' = nL$. 
Therefore 
\begin{equation}
\label{eq:inflcfl}
f_l(x') = -k \sum_{n=-\infty}^\infty \Delta_n(x') (nL-x')
\end{equation}
where $\Delta_n(x')$ is the probability of initially finding the chain end
between $nL$ and $(n+1)L$, 
\begin{equation}
\Delta_n(x) = E(nL-x)-E((n+1)L-x).
\end{equation}
It is easily seen that $\Delta_{-n}(L-x) = \Delta_n(x)$. Using this and Eq. \ref{eq:inflcfl}
\begin{equation}
f_l(L-x) = -k(-f_l(x) -L \sum_{n=-\infty}^\infty \Delta_n(x')) =  -f_l(x) +kL  .
\end{equation}
To obtain the work, we follow the same procedure as above and consider $f_l(x') + f_l(L-x')$, which
here is just $kL$. Therefore in this case,
\begin{equation}
W_L = \frac{c_0}{c_0+c_1} \frac{k L^2}{2}
\end{equation}
which in this model exactly, for all $k$ and $L$, given that we are in a region
of strong metastability.

\subsection{Efficiency in large power stroke limit}
\label{subsec:EffLargePowerStroke}

We now are in a position to answer a central question about the performance of
this kind of device: is the efficiency limited by the small value of $k_B T$
compared to photon energies? We have seen from the above analysis that it is
possible to get arbitrarily large forces developing at the expense of
exponentially slow operation. However this is in the limit of infinitesimal
plate velocity $\rm v$. In contrast, the power obtained is instead average
force times this, $\langle f\rangle {\rm v}$ and we would like to operate the
device at the velocity of maximum power, which necessitates the operation of it
far from equilibrium, because compensating the increase in $\rm v$ is a
decrease in $\langle f\rangle$ due to dissipation. It could be that the optimum velocity of operation
decreases very quickly with $k L^2/2$ meaning that the device becomes
increasingly inefficient as the spring constant $k$ is increased. This would
make it impossible to harvest more than of order $k_B T$ energy per cycle.

The efficiency is defined as the ratio of the amount of power produced to the
amount of power put in. The amount of energy needed to dissociate a polymer end
from a binding site is $V_{min}$. Binding to $V_{min}$ must produce an energy less
than that of an unbound polymer. The power put in is $V_{min} c$ where $1/c =
1/c_0 + 1/c_1$. Therefore the efficiency is
\begin{equation}
\label{eq:efficiency}
\eta = \frac{\langle f\rangle {\rm v}}{c V_{min}} .
\end{equation}

To investigate this problem further, the one dimensional model of the last
section with $l_c\rightarrow \infty$ was implemented using Metropolis Monte Carlo.
We chose the periodic potential $V_p$ to vary as
\begin{equation}
\label{eq:Vp1dsim}
V_p(x)  =\begin{cases}
-4 V_{min} \frac{x}{\delta}(1-\frac{x}{\delta}),  & \text{if $0 < x< \delta$}.\\
-4 V_{max} \frac{x}{\delta}(1+\frac{x}{\delta}),  & \text{if $0 > x> -\delta$}.\\
0, & \text{otherwise}.
\end{cases}
\end{equation}
Here we set $\delta = 0.1$, $V_{max} = V_{min} = 100$, and $L=1$. 
In order to preserve diffusional dynamics, steps in x were attempted uniformly 
in the range $[-0.025,0.025]$ ensuring that the periodic potential cannot
be jumped across in one move. One move increased the time by $.05$, though this
number was arbitrary and aside from an obvious rescaling, does not affect the results obtained.

To observe behavior in the limit of metastability as discussed in the previous
section, the rate of unbinding must be set to be small compared to the inverse
metastable equilibration time in the bound state. Hence we chose the unbinding
rate $c_0 = 10^{-5}$ and $c_1 = 5\times 10^{-4}$. 

\begin{figure}[htp]
\begin{center}
(a)\includegraphics[width=\hsize]{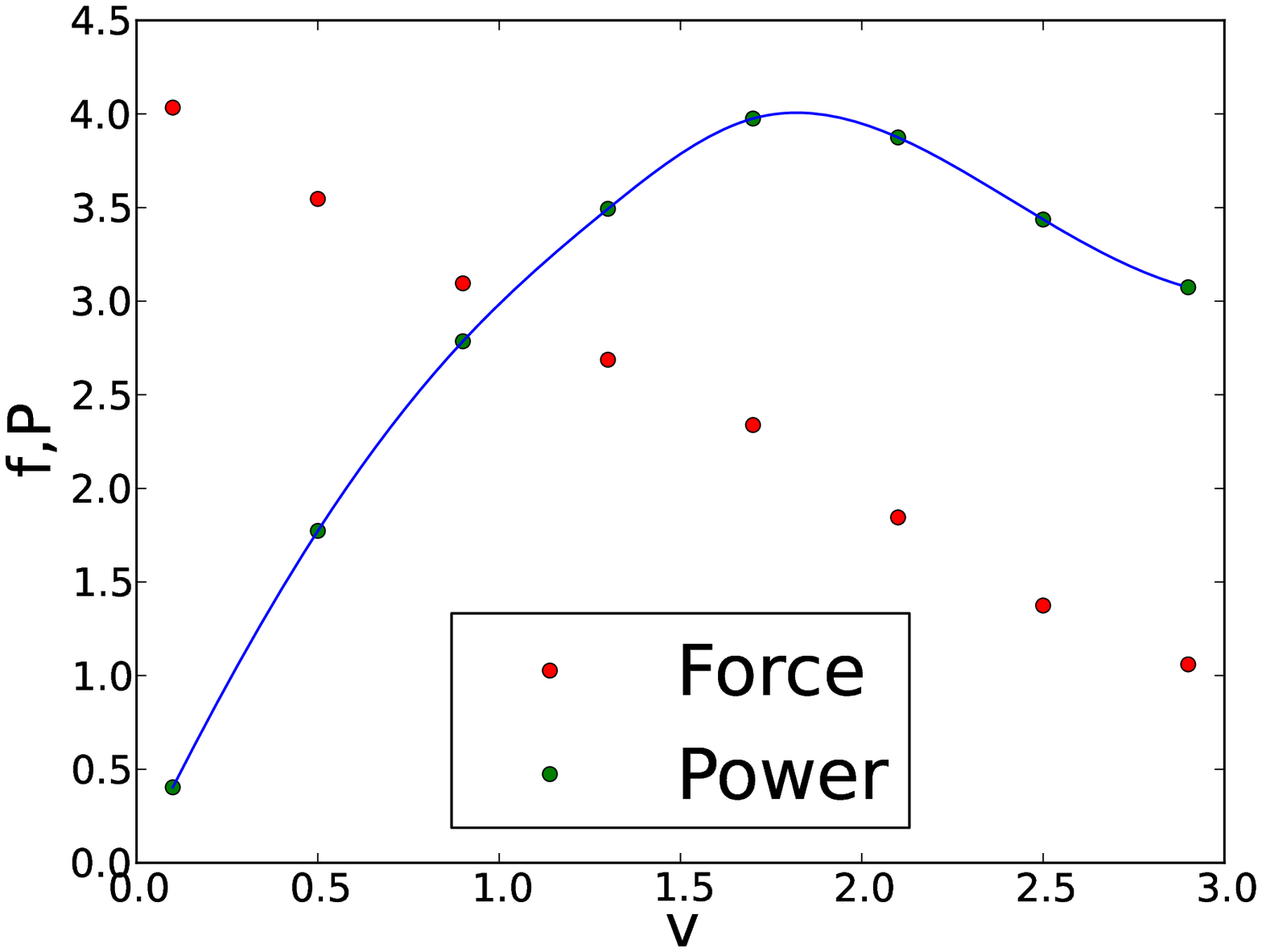}\\
(b)\includegraphics[width=\hsize]{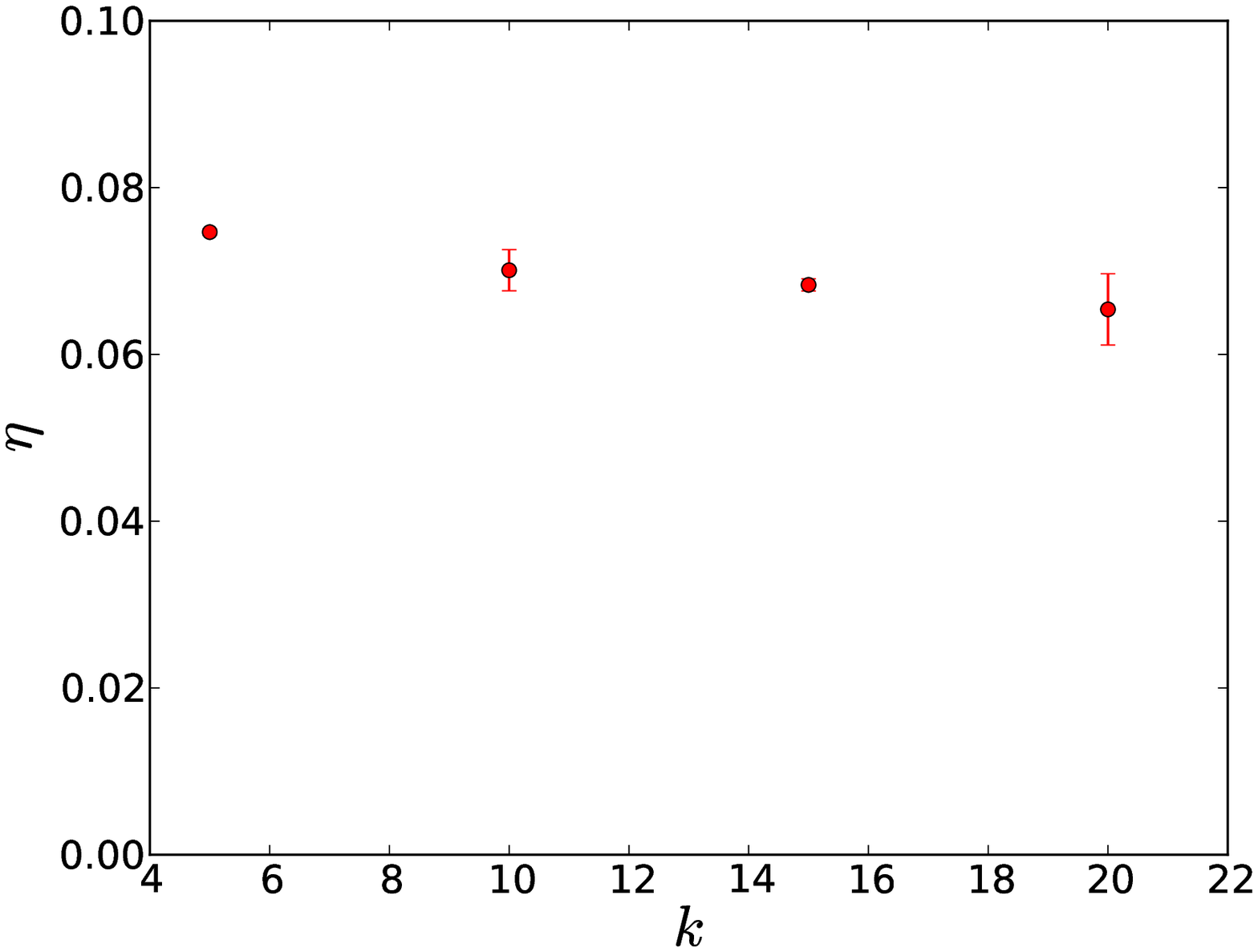}
\caption{
(a) The force and power versus velocity for the one dimensional model described
in the text. Here the spring coefficient $k=15$. The line going through the
power is a cubic spline fit used to more accurately determine the maximum value
of the power. The velocity and power are both in units of $10^{-6}$.
(b) A plot of the efficiency of the motor as a function of the spring constant
$k$. Three separate runs, each of $3\times 10^{10}$ steps were used to determine
the error bars, for each data point shown.
}
\label{fig:1dfvsv}
\end{center}
\end{figure}

As the simulation was running, the tether point was moved at velocity $\rm v$
which was typically small, 
The average spring force $\langle f\rangle$ was measured as a function of $\rm v$
for a given spring constant $k$. The results of a run of $3\times 10^{10}$ steps
are shown in Fig. \ref{fig:1dfvsv}(a).

We are interested in the limit of large $k$ although this is hard to achieve
numerically owing to exponentially long relaxation times. The parameters used
allow us to probe up to $k=20$. 
In the large $k$ limit, the  minimum value of $V_{min}$ needed to bind, is
$k l^2 /2$, where $l$ is the maximum amount the spring will need to stretch
from the tether to the binding site. Because $\delta = 0.1$ and $L=1$, this
implies $l=0.9$. Therefore our formula for the efficiency, given this input
energy is $\eta = P_{max}/(c k l^2 /2)$, where $P_{max}$ is the maximum of the
power versus velocity, as shown in Fig. \ref{fig:1dfvsv}(a). Plotting the
efficiency for different values of $k$ ranging from $5$ to $20$ yields the
points in Fig. \ref{fig:1dfvsv}(b). 
Despite the fact that the relaxation time for metastable relaxation of the system
varies over more than two orders of magnitude, the overall efficiency is almost
constant. We expect at higher value of $k$, the efficiency will eventually drop
owing to the fact that the average unbinding time becoming smaller than the
metastable relaxation time.

What the above analysis shows is that in the limit where the photon cross section can be made arbitrarily low, the efficiency 
can be adjusted to be constant, independent of the photon energy. This is
accomplished by choosing a large spring coefficient. In reality
with photon energy of $2 eV$ and $k_B T \approx 1/40 eV$, photon flux would have
to be far too low for this optimal regime to be realizable. However the
above analysis also shows that the efficiency can be substantially increased by
choosing larger $k$ at the expense of lowering the cross section. The energy
delivered in one cycle should scale as $E_c = k L^2$. This can be increased to
be substantially larger than thermal energies but at the expense of a long
relaxation time proportional to $\exp(E_c/k_B T)$. As noted earlier, it should
be possible to make $E_c$ about $7 k_B T$, with reasonable parameter estimates. 

\section{Conclusions}
\label{sec:Conclusions}

Here we have analyzed the viability of converting photons to mechanical
energy using a device composed of an canted polymer brush tethered to a
lower plate but able to bind its other ends to sites on an upper plate. Photons can dissociate
these ends from binding sites. By a combination of analytical and numerical arguments we showed
that in steady state, this produces net mechanical power.

The system is inspired by biological motors such as myosin II that bind
to actin and is dissociated by the binding of ATP. The analysis used here
could also be applied to such systems, however in reality they contain
many more stages. In general, these kind of systems are classified as 
``thermal ratchets"~\cite{ReimannRev}, where the system can be thought of as moving in a
washboard potential in the presence of thermal noise. Though that description
can be very useful in understanding the general principles behind
the operation of such motors, in the present case, we are trying to model
the system in more detail than such models can afford. Instead we have described
the system using Langevin dynamics and also two coupled Fokker-Planck equations similar
in spirit but not identical to previous approaches~\cite{ProstPRL,BustamanteKellerOster}. The difference
here is that for potentials to be a sensible model for a polymer tethered
to a single point, they cannot be periodic. Such modelling allows us to
see how varying microscopic parameters affect the power and force
characteristics.

The analytical results on the Unbound Equilibration Model and extensive one
dimensional simulations, show that the force
applied by the device can be made arbitrarily large at the expense of having
exponentially long relaxation times. At a given photon flux, the production of large forces 
from single polymers imply the need for a very low 
cross section of interaction between the photon and the bound end plus binding
site, as long relaxation times are required. Therefore there is a trade off between this force
and the speed the device can move. This may be circumvented to some extent by
stacking transparent devices of this kind, so that even though the cross-section of interaction of an
individual photon with a given layer is low, it will eventually be captured
by one layer. 

Therefore it appears that there is no theoretical obstacle to prevent
the photo-mechanical conversion of energy in this manner, however
it represents a significant experimental challenge.

It has been pointed out~\cite{GrzybowskiReview} that there is an important
distinction between artificial molecular switches and artificial molecular
machines, the former being a fraction of a penny, and the latter being extremely
challenging to create. The approach investigated here is closer to biological
motors than other proposals, and should be more forgiving about randomness, either 
during fabrication, or due to thermal motion, than approaches that require precise chemical synthesis
of molecules capable of sophisticated conformational changes~\cite{credi2006}.
However its experimental realization is still quite clearly a formidable task.

\section{Acknowledgments}

The author would like to thank Professor Monica Olvera de la Cruz and Edward
Santos for useful discussions. Support from  National Science Foundation
CCLI Grant DUE-0942207 is gratefully acknowledged.

{}
\end{document}